\documentclass[letterpaper, 10 pt, conference]{ieeeconf}
\IEEEoverridecommandlockouts
\makeatletter
\def\endthebibliography{%
  \def\@noitemerr{\@latex@warning{Empty `thebibliography' environment}}%
  \endlist
}
\makeatother

\usepackage{graphicx}      

\PassOptionsToPackage{hyphens}{url}\usepackage{hyperref}

\usepackage{amsmath} 
\usepackage{siunitx}

\usepackage{amsthm}
\usepackage{mathtools}
\usepackage{verbatim} 
\usepackage{soul}
\usepackage[linesnumbered,ruled]{algorithm2e}
\usepackage{bm}
\usepackage{subfigure}
\usepackage[export]{adjustbox}
\usepackage{color}
\usepackage{xcolor}
\usepackage{tikz}
\let\labelindent\relax
\usepackage{enumitem}
\usepackage{epsfig} 
\usepackage{relsize}
\usepackage{booktabs}
\usepackage{multirow}
\usepackage[super]{nth}

\def\IEEEsetlabelwidth#1{\settowidth{\labelwidth}{#1}}
\def\IEEEusemathlabelsep{\setlength{\labelsep}{\IEEEiedmathlabelsep}}

\makeatletter
\newcommand\xmathlarger[2][1]{%
  \mbox{\larger[#1]$\displaystyle#2\m@th$}%
}
\makeatother

\usepackage{url}
\usepackage{soul,color}
\usepackage{balance}

\theoremstyle{definition} 

\begin{document}
\title{Regulating EV Charging Markets for Fairness: Incentives for Pricing and Capacity Decisions}

\author{Ruiting Wang, Kita Hu, Yitong Yu, Scott Moura 
\thanks{This work was supported by Katherine S. and James K. Lau Graduate Fellowship in Climate Equity. }
\thanks{R. Wang, and S. Moura are with the Department of Civil and Environmental Engineering, University of California, Berkeley, CA, 94720 USA. \tt smoura@berkeley.edu} 
\thanks{R. Wang is also with the School of Electrical Engineering and Computer Science, KTH Royal Institute of Technology, Stockholm, Sweden.  \textit{Corresponding author: Ruiting Wang} \tt ruiting@kth.se}
\thanks{K. Hu is with the Department of Mathematics, University of California, Berkeley, CA, 94720 USA. \tt kita\_0314@berkeley.edu}
\thanks{Y. Yu is with School of Civil and Environmental Engineering, Nanyang Technological University, 50 Nanyang Avenue, Singapore 639798, Singapore. \tt yitong001@e.ntu.edu.sg}
}

\maketitle
\begin{abstract}
The transition to electric mobility calls for charging infrastructure that is both efficient and socially equitable. This paper examines fairness in electric vehicle (EV) charging station pricing and capacity through a game-theoretic perspective. We model a non-cooperative market in which competing charging service providers set prices and capacities while customers choose stations based on generalized cost, leading to a market equilibrium. We then benchmark this decentralized outcome against an idealized planner solution that jointly optimizes efficiency and equity. To align market outcomes with socially desirable goals, we design targeted incentives that guide operators toward more fair charger placement. Case studies demonstrate that unregulated competition tends to exacerbate disparities in charger access across demographic groups, whereas carefully calibrated incentives can reduce inequities without significant efficiency loss. The framework provides insights for policymakers on reconciling free-market dynamics with the broader societal goals of fairness in electrified mobility systems.
\end{abstract}

\section*{Nomenclature}

\subsection{Parameters}
\begin{IEEEdescription}[\IEEEusemathlabelsep\IEEEsetlabelwidth{$K$}]
    \item[$G_j$] Customer Generalized Cost Function for Station $j$.
    \item[$K_j$] Daily capital cost investment per unit capacity for chargers at Station $j$.
    \item[$N$] Total number of customers in the market.
    \item[$P_j$] Daily energy purchase cost per unit of customer at Station $j$.
    \item[$T_{j,0}$] Free-flow travel and charge time for Station $j$.
    \item[$o^0_j$] Benchmark opportunity cost outcome for Station $j$.
    \item[$w$] Weighing parameter for fairness in Global Opt.
    \item[$\alpha_j$] Congestion sensitivity coefficient for Station $j$. 
    \item[$\lambda$] Sensitivity parameter for customer choice (in the logit model).
    \item[$\tau$] Value of time (monetary equivalent per unit of time).
    \item[$\pi$] Profit Function for Station $j$.

\end{IEEEdescription}

\subsection{Variables}
\begin{IEEEdescription}[\IEEEusemathlabelsep\IEEEsetlabelwidth{$K$}]
    \item[$p_j$] Monetary price charged by Station $j$, assuming all customers are charged the same amount of money. 
    \item[$c_j$] The aggregate charging capacity at Station $j$.
    \item[$q_j$] Congestion level (number of customers) in Station $j$.
    \item[$T_j$] Additional time needed for charging at Station $j$.
    \item[$o_j$] Opportunity cost of customers visiting Station $j$ by the cumulative time cost.
\end{IEEEdescription}

\section{Introduction}
\subsection{Background}

Electric car sales hit 17 million worldwide in 2024 \cite{IEA2025}. The rapid growth of EV adoption places increasing strain on the charging infrastructure, making the efficiency and equity of electric vehicle charging a pressing engineering and policy challenge. In many regions, charging stations operate in a competitive market environment. Several major Charge Point Operators (CPOs), such as Tesla, Ionity, and ChargePoint, independently set pricing and capacity \cite{BCG2025}. This setting naturally lends itself to a game-theoretic perspective, where stations compete for customers. The resulting equilibrium reflects the dynamics of individual station decisions and customer charging behavior. However, increasing attention has been given to how this profit-driven market may lead to an inequitable distribution of charging resources, leaving certain disadvantaged communities behind in the global process of electrification. 
This research examines how policy interventions can help balance equity and efficiency in a decentralized market by steering the market equilibrium toward a globally preferred outcome.

\subsection{Decentralized Charging Market}

Several studies applied game-theoretic models to capture the interaction between CPOs and customers, typically as a leader–follower Stackelberg game. One early study examined a smart grid coordinating with multiple groups of EVs, showing that the followers’ non-cooperative behavior can reach a generalized Nash equilibrium when electricity capacity is fixed \cite{tushar2012}. Another work focuses on two competing stations receiving Poisson-arriving EVs: prices are announced first by the stations, then drivers select where to charge, and an equilibrium exists that depends on the price differential \cite{yuan2017}. 

Customers are commonly modeled as cost minimizers, distributing their route choice across charging stations, which leads to user equilibrium according to Wardrop's First Principle \cite{correa2011}. 
In existing literature, the customer’s cost function often includes charging price and travel distance, but time is inconsistently modeled. Examples include a time sensitivity factor \cite{sun2025}, waiting time \cite{yuan2017} with Poisson arrivals, or travel time is ignored in the utility function \cite{lee2015}. 

Given the existence of a Nash equilibrium among companies and travelers, recent work has shifted toward bilevel models to connect leaders and followers with a Stackelberg equilibrium to maximize company profit and minimize consumer cost. Yang \textit{et al.} \cite{yang2022} generalized this concept in the competitive carsharing market, proposing a solution to the multi-leader common-follower game where operators optimize pricing and relocation strategies while customer minimizes disutility. Sun \textit{et al.} \cite{sun2025} proposed a bilevel planning model with Nash bargaining game, balancing operators’ investment costs with EV users’ queuing time and satisfaction; their model integrates demand forecasting with pile-type preferences, reducing both user waiting time and system cost. Zeng \textit{et al.} \cite{zeng2020} used a similar bilevel model to capture strategic decisions made by EV users in the lower level and optimize for the best charging station practice and pricing scheme in the upper level. Their optimization formulation aimed to account for uncertainty in wholesale energy prices, renewable-source availability, and EV flow.

\subsection{Equitable Infrastructure Design}

Nowadays, inequitable access to EVCSs remains a critical barrier to widespread adoption of EVs. 
Recent work showed that, high-income communities, often comprising single-family homeowners with built-in EV chargers, enjoy easier access to chargers and have a higher EV adoption rate; whereas low-income communities with limited access to EVCS have lower adoption rates due to higher perceived cost \cite{hsu2021a}. 
As CPOs are motivated by profitability, ECVS installation would be in higher-adoption areas to ensure maximum utility. This pattern reinforces inequitable charging in underserved communities \cite{wang2026}. 

Much literature has explored the charging resource accessibility in recent years, by the spatial coverage of charging resources \cite{hanig2025}, by the visit-based approach using mobility patterns from cell phone data \cite{qian2025a}, by charging opportunity \cite{kontou2019}, and potentially more. Meanwhile, regional case studies reported race/ethnicity, income, and urban/suburban as the most prominent imbalance in equitable charger access \cite{khan_inequitable_2022}, \cite{carlton_electric_2022}, \cite{roy_examining_2022}. These observations highlighted insufficient exploration of equity considerations in charger placement and affordability, which calls for more studies on equitable considerations for EVCS placement and pricing. However, none of the equity-related papers consider this problem from a market perspective, where CPOs' planning and operation are mainly profit-driven. 



\subsection{Contribution}

While infrastructure planning ought to harmonize equitable access with operational efficiency, the current landscape, dominated by private CPOs, favors profit-driven allocation. This shift is resulting in inequitable charging access, even service deserts in some communities. 
We introduce a game-theoretic bi-level model that captures the market's competitive equilibrium, define a socially optimal state incorporating fairness, and design an incentive structure to bridge the difference. Our approach provides a practical framework for policymakers to reduce socioeconomic disparities in the oligopolistic EV charging market without significantly compromising operational efficiency.


\section{Decentralized Market Equilibrium Model}

Consider a single origin-destination (OD) pair with $N$ travelers and $M$ competing stations, indexed by $j \in \{1, 2, \dots, M\}$, serving as potential charging options \footnote{For a city with multiple OD pairs, each pair of ODs will be analyzed independently, but congestion levels at each station introduce a coupling between these separate problems.}.

\subsection{Lower-Level Equilibrium: Customers}

Each customer chooses a station with a probability given by a Multinomial Logit (MNL) model, which conceptualizes the decision-making process as a discrete choice driven by random utility maximization.
This model acknowledges that while agents seek to minimize generalized costs (in the following section), their choices are influenced by unobserved preference heterogeneity and bounded rationality.

Given the generalized cost for a customer choosing Station $j$ to be $G_j$, the total customer demand $N$ is distributed among the $M$ stations according to the MNL framework. The flow $q_j$ to Station $j$ is:

\begin{equation}
    q_j = N \cdot \frac{\exp(-\lambda \cdot G_j)}{\sum_{k=1}^{M} \exp(-\lambda \cdot G_k)} \label{eq:flow_allocation}
\end{equation}

We consider the generalized cost for a customer choosing Station $j$ is a function of its price, travel time, and station congestion level: 

\begin{equation}
    G_j(p_j, q_j, c_j) = p_j + \tau \cdot T_j(q_j, c_j) \label{eq:generalized_cost}
\end{equation}
where $T_j(q_j, c_j)$ is the time needed for charging at Station $j$. This time consists of a free-flow detour duration, and a congestion term, the latter increases with customer demand $q_j$ and decreases with capacity level $c_j$. We here use a linear simplification for $T_j$:

\begin{equation}
T_j(q_j, c_j) = T_{j,0} + \alpha_j \cdot \frac{\max(0, q_j - c_j)}{c_j} \label{eq:travel_time_revised}
\end{equation}

This formulation constitutes a fixed-point problem, as the generalized cost is endogenously determined by the prevailing congestion levels, which is a realization of the aggregated MNL choices. 


\subsection{Upper-Level Equilibrium: Stations}

Each Station $j$ acts as a rational economic agent, choosing its price $p_j$ and capacity $c_j$ to maximize its own profit, anticipating the customers' response (i.e., how demand allocates with customer choice $q_j$ derived from the lower-level problem). This forms a Nash Equilibrium in an oligopolistic market.

The profit for Station $j$ is its revenue minus its fixed and operating costs.

\begin{equation}
\pi_j = q_j \left( p_j - P_j - \frac{\tau \cdot \max(0, q_j - c_j)}{c_j} \right) - K_j c_j
\label{eq:profit_function_revised}
\end{equation}

where:
\begin{itemize}
    
\item $K_j$: Fix cost initial investment of providing per unit of charger capacity. 
\item $P_j $: Cost associated with purchasing per unit energy to serve customers.
\item $\frac{\tau \cdot\max(0, q_j - c_j)}{c_j}$: Monetary penalty for congestion time. 
\end{itemize}

Each station $j$ solves the following optimization problem to maximize its profit, taking the strategies of all other stations $k \neq j$ as given:
\begin{equation}
    \max_{p_j, c_j} \pi_j \label{eq:station_optimization}
\end{equation}
subject to:
\begin{align*}
    p_j &\ge 0 \quad \text{(Prices must be non-negative)} \\
    c_j &\ge 0 \quad \text{(Number of chargers must be non-negative)}
\end{align*}

This entire problem is a form of Non-Linear Program (NLP). Specifically, it is a bilevel program where the upper-level problem (stations' profit maximization) depends on the solution of the lower-level problem (customers' charging choice).

\section{Centralized Planner}

In this part of the model, we consider an alternative scenario in which the charging infrastructure is centrally owned and managed by a government planner who seeks to optimize system efficiency while also ensuring equity in the distribution of charging resources.
\subsection{Utilitarian Perspective}
As a benchmark, we examine the case where the central planner operates purely from a utilitarian perspective, minimizing the cumulative time cost of all customers. 
We assume that the opportunity cost of customers' visiting is captured solely by the cumulative time cost $o_j = q_j \tau T_j(q_j, c_j)$. 
In this scenario, the planner seeks to minimize the total system cost, expressed as, 

\begin{equation}
     \min \sum_j o_j
\end{equation}

The system is subject to constraints \eqref{eq:flow_allocation}-\eqref{eq:travel_time_revised}. Additionally, an overall capacity budget is imposed to ensure the system remains bounded. For consistency, we set the total capacity equal to that of the decentralized market results.

\subsection{Fairness Perspective}
Meanwhile, the planner’s objective is not limited to minimizing the total system cost, but also to balancing efficiency with distributional equity. An idealized allocation of charging resources should therefore account for both opportunity cost and fair accessibility across different user groups. System performance is evaluated along two dimensions:

\begin{enumerate}
    \item Efficiency: measured by the aggregate opportunity cost across all stations, $\sum_j o_j$. 
    \item Equity: measured by how different groups benefit from the allocation relative to a benchmark scenario.
\end{enumerate}

To capture fairness, our formulation is inspired by the Nash Bargaining Solution (NBS), which evaluates allocations based on opportunity improvements relative to a benchmark. In this framework, equity is promoted by maximizing the product of group-level improvements, which discourages allocations where one group obtains disproportionately large gains while others benefit only marginally, thereby yielding more balanced outcomes.
This approach sits naturally between pure utilitarian efficiency (which may sacrifice disadvantaged groups) and Rawlsian fairness (which prioritizes the worst-off group exclusively) \cite{hall_limits_2025}. Formally, the NBS gain is written as
\begin{equation}
O_{\mathrm{NBS}} \;=\; \prod_{j} \bigl(o^0_j - o_j\bigr)
\end{equation}
which measures the joint surplus over the benchmark of the allocation $o^0_j$.
For numerical stability and to facilitate optimization, we equivalently maximize the logarithm of the objective, or equivalently, minimize its negative:

\begin{equation}
\min_{c} - \sum_j \log \bigl(o^0_j -  q_j \cdot \tau \cdot T_j(q_j, c_j) \bigr)
\end{equation}

Note that the domain requires $o^0_j > o_j$ for all $j$. The planner thereby seeks an allocation that jointly balances system efficiency and group-level equity.

\section{Methods}

\subsection{Algorithm for Decentralized Market}
The solution approach combines three algorithmic components: the Method of Successive Averages (MSA) for computing user equilibrium flows, a best-response optimization procedure for operator profit maximization, and an iterative Nash equilibrium search across competing companies.  

\subsubsection{Method of Successive Averages}  
The MSA algorithm is used to determine user equilibrium flows given station prices and capacities \cite{powell1982}. Since customer flows appear implicitly in both generalized cost functions and probabilistic choice models, the problem is formulated as a fixed-point system. MSA resolves this system through successive averaging.
After an initialization step, at each iteration $s$, auxiliary flows $q^{aux}$ are computed based on the current generalized costs and the multinomial logit model \eqref{eq:flow_allocation}. The flows are then updated using the rule:
\begin{equation}
   q^{(s+1)} = q^{(s)} + \frac{1}{s}\big(q^{aux} - q^{(s)}\big), 
\end{equation}
which gradually adjusts the solution toward equilibrium. The procedure terminates once the difference between consecutive flow updates falls below a predefined tolerance, ensuring convergence to a stochastic user equilibrium.

\subsubsection{Best-Response Optimization}  
Given equilibrium flows obtained from MSA, each company updates its operational strategy to maximize profit. This procedure takes into account both pricing and capacity investment decisions. We relax the optimization to a continuous formulation and solve it using the L-BFGS-B algorithm \cite{zhuciyou1997}, a quasi-Newton method tailored for smooth nonlinear problems with simple bound constraints, to efficiently compute each station’s best-response. Once the decision variables are obtained, station capacities are rounded to integers, and the solution is checked for convergence.

\subsubsection{Diagonalization Algorithm}  
The interaction among competing companies is modeled as a non-cooperative game.  
To capture this interaction, we adopt a diagonalization algorithm, a widely used approach for computing Nash equilibria in network and energy markets. The algorithm proceeds iteratively: in each round, user flows are first determined via the MSA procedure, ensuring consistency with stochastic user equilibrium. Given these flows, each company then solves its best-response optimization problem while treating the strategies (prices and capacities) of its competitors as fixed. The resulting updates in prices and capacities are applied sequentially across all companies.  
This iterative process continues until convergence, defined as the state where no company can increase its profit by unilaterally deviating from its current strategy. At this point, the system has reached a Nash equilibrium, representing a stable outcome of the competition.

The overall solution procedure can be summarized in the following pseudo-code:

\begin{algorithm}[h]
\SetAlgoLined
\KwIn{Initial station strategies $\{(p_j^0, c_j^0)\}$, total demand $N$, logit parameter $\lambda$, travel cost coefficient $\tau$, congestion sensitivity $\alpha$, tolerances $\epsilon_p, \epsilon_c$, maximum iterations $T$}
\KwOut{Nash equilibrium strategies $\{(p_j^*, c_j^*)\}$ and flows $\{q_j^*\}$}

\For{$t = 1$ \KwTo $T$}{
    \tcp{Compute eq. flows via MSA}
    compute auxiliary flows $q_j^{aux} = N \frac{\exp(-\lambda G_j)}{\sum_k \exp(-\lambda G_k)}$\;
    update flows $q_j^{(t+1)} = q_j^{(t)} + \frac{1}{t} (q_j^{aux} - q_j^{(t)})$\;

    \For{each station $j$}{
        \tcp{Joint best-response optimization for station $j$}
        $(p_j^{new}, c_j^{new}) = \arg\max_{p,c} \text{Profit}_j(p,c;\{q_k^{(t+1)}\})$\;
        update station strategy: $p_j^{(t+1)} \gets p_j^{new}, \quad c_j^{(t+1)} \gets \text{round}(c_j^{new})$\;
    }

    compute maximum changes: $\Delta_p = \max_j |p_j^{(t+1)} - p_j^{(t)}|$, $\Delta_c = \max_j |c_j^{(t+1)} - c_j^{(t)}|$\;
    
    \If{$\Delta_p < \epsilon_p$ \textbf{and} $\Delta_c \le \epsilon_c$}{
        break\tcp{Nash equilibrium reached}
    }
}

compute final flows $\{q_j^*\}$ via MSA using equilibrium strategies\;
\Return{$\{(p_j^*, c_j^*)\}, \{q_j^*\}$}
\caption{Nash Equilibrium Computation for Competitive EV Charging Stations}
\end{algorithm}

The final outcome represents a nested equilibrium structure: user-level SUE embedded within operator-level Nash equilibrium in pricing and capacity decisions.  


\subsection{Algorithm for Centralized Market}  

The centralized optimization problem is solved using the commercial solver Gurobi (version 12.0.3)\cite{gurobi}.

\subsection{Experimental Setting}  

We consider a total of $N = 100$ customers traveling through the same OD pair, and can choose from $M = 4$ competing stations. Each station requires some detour from the main road, which is captured by $T_{j,0}$. The rest of the parameters are summarized in Table~\ref{tab:exp_params}. In particular, we use initial utility values in the original market equilibrium as the benchmark $o^0_j$ used in the NBS perspective. 
The code for this work can be found at: \url{https://github.com/TIVV424/EV-charge-game}.

\begin{table}[h]
\centering
\caption{Experimental parameters}
\label{tab:exp_params}
\begin{tabular}{l c p{0.55\linewidth}}
\toprule
\textbf{Parameter} & \textbf{Value} & \textbf{Meaning} \\
\midrule
$N$ & 100 & Total demand (number of charging requests). \\
$M$ & 4 & Number of candidate charging stations (same number of competing firms). \\
$T_{j,0}$ & [10, 7, 5, 3] & Detour time to reach each station from main road.\\
$\lambda$ & 0.6 & User sensitivity to generalized cost in the logit model. \\
$\tau$ & 0.5 & Weight of travel time in the generalized cost. \\
$\alpha$ & 20 & Congestion parameter influencing queuing effects [min]. \\
$K_j$ & 15 & Fix cost rate per unit of charging capacity [USD/char]. \\
$P_j$ & 5 & Operating cost rate for serving one customer [USD/cus]. \\
\bottomrule
\end{tabular}
\end{table}  

\section{Results}

\subsection{Decentralized versus Centralized Outcomes}

We compare three scenarios: a decentralized market equilibrium, a utilitarian centralized optimization, and a centralized optimization incorporating fairness considerations. To ensure a meaningful comparison, the total capacity in both centralized scenarios is constrained to match the capacity resulting from the decentralized equilibrium.

\begin{figure*}[hbtp]
    \centering
    \includegraphics[width=\linewidth]{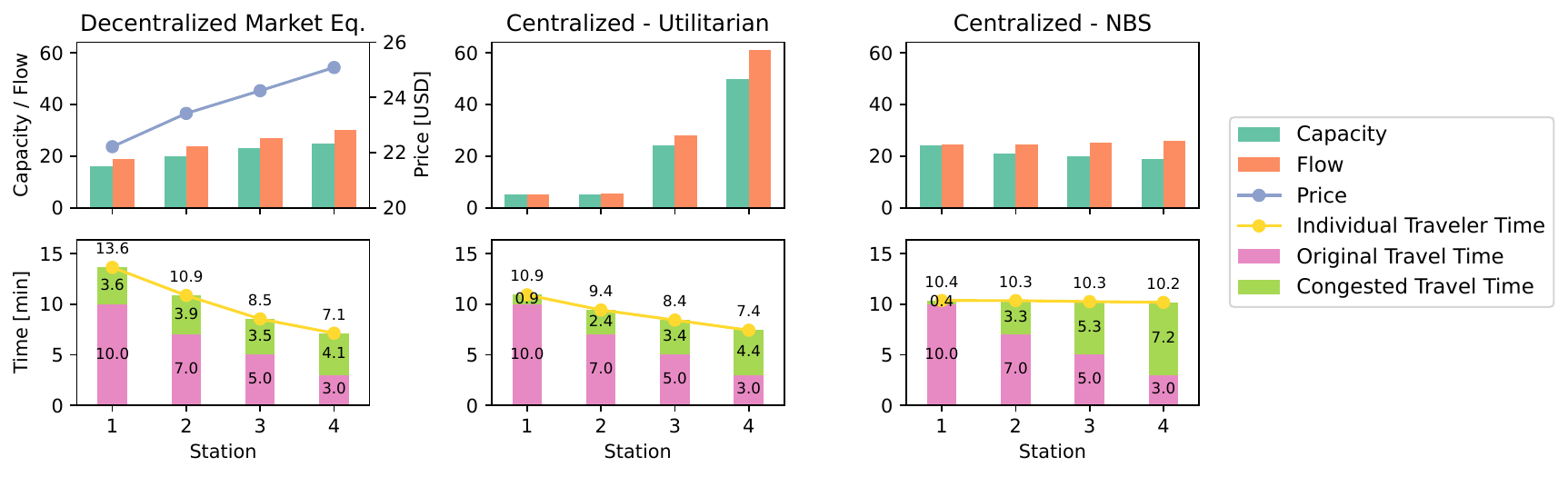}
    \caption{Comparison of three scenarios: decentralized market equilibrium, centralized optimization from a utilitarian perspective, and centralized optimization incorporating fairness via the Nash Bargaining Solution. For each scenario, we plot planned capacities and flows, with total capacity held constant across all cases. The time plots show individual customer travel time, broken down into original travel time and additional congested time.}
    \label{fig:result1}
\end{figure*}

For each scenario, we plot the planned capacities and flows and illustrate the two components of user travel time: the initial detour time to the station and the additional congestion delay (i.e., waiting time), as shown in Fig. \ref{fig:result1}.

In the decentralized market, each station operator, with knowledge of customer behavior, strategically adjusts capacity and avoids oversizing to maintain a certain level of congestion. This allows popular stations to exploit high demand by setting higher prices to maximize profit. Conversely, a subset of less time-sensitive customers opt to travel farther distances to benefit from lower charging prices at less congested stations. 

The second panel reveals a key limitation of the centralized, utilitarian optimization. From this perspective, the system prioritizes installing chargers at the most accessible location (Station 4), as indicated by its low initial travel time (pink bars), to minimize total system delay. However, this ``optimal'' solution overlooks critical real-world complexities. First, such centrally located stations often have practical constraints, such as limited land for expansion. Second, from a network perspective, focusing on the closest station to one main road creates a local optimum that poorly serves the broader system. A slightly more distant station might be better positioned to serve multiple traffic corridors, leading to a more efficient and resilient global optimum. Consequently, the utilitarian approach risks creating bottlenecks and inequitable access, underscoring the need for models that incorporate fairness.

The third panel, illustrating the centralized optimization with fairness considerations, achieves a more balanced distribution of user time costs across different stations. 
This example captures only a limited dimension of the broader fairness challenge, namely ensuring that users experience similar total time costs regardless of the station they select.
In this formulation, fairness is promoted by reducing the extent to which outcomes are determined by initial travel distances.

While the utilitarian optimization expectedly\footnote{This follows directly from the objective function used.} achieves the most efficient outcome with the lowest total travel time (799.9 min), incorporating fairness introduces a clear trade-off. The fairness-oriented scenario increases the total time (1028.3 min) as a ``cost of fairness'' required to achieve a more equitable distribution of charging-related time across users.

\subsection{Market Response with Intervention}
To align the decentralized market outcome more closely with the fairness-oriented allocation, we introduce a subsidy on per-unit charging capacity. We use an iterative algorithm to compute per-unit capacity subsidies, with the goal of guiding the system of EV charging stations toward equitable travel times across stations. This can be viewed as a searching process at the regulatory level: subsidies are iteratively adjusted until stations’ endogenous choices (prices, capacities) and users’ responses converge to a balanced outcome with aligned travel times. Since it effectively adds an outer layer on top of the Nash equilibrium computation, and due to space constraints, we do not elaborate on it as a separate algorithm.

\begin{figure}[hbtp]
    \centering
    \includegraphics[width=\linewidth]{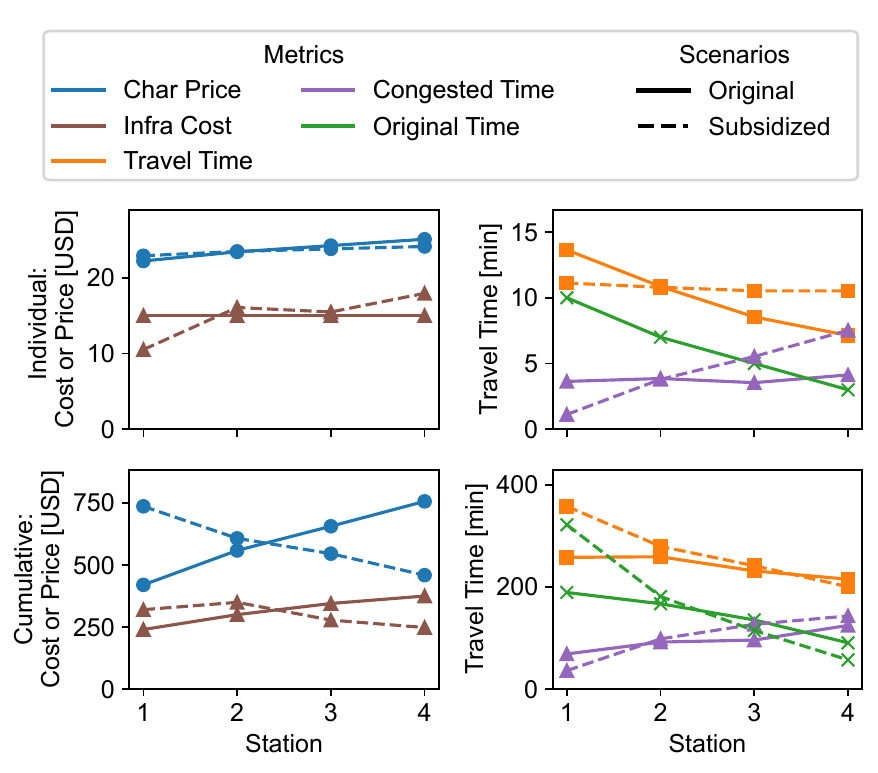}
    \caption{Comparison of decentralized market outcomes for baseline (solid lines) and subsidized (dashed lines) scenarios. The top panels illustrate individual-level metrics (price, cost, and time spent), while the bottom panels show aggregate, station-level metrics, including traffic flow and cumulative time delay. Performance metrics are differentiated by color.}
    \label{fig:subsidy}
\end{figure}

This intervention alters the investment incentives of charging station operators by reducing the effective cost of capacity expansion. As a result, operators are more willing to allocate capacity to stations that are less favored in the purely profit-driven equilibrium. The result is shown in Fig. \ref{fig:subsidy}.

The subsidy reshapes the market response in two important ways. First, it mitigates the excessive congestion observed at the most accessible stations by encouraging additional capacity investment at more distant stations. Second, it redistributes customer flows in a manner that reduces disparities in travel time across groups. Consequently, the resulting allocation converges toward the NBS, which balances efficiency and equity by improving proportional outcomes across all groups relative to their benchmarks. 

Overall, the intervention demonstrates that relatively simple economic instruments, a per-unit capacity subsidy, can effectively steer decentralized market behavior toward socially desirable outcomes without requiring full centralization.

\section{Conclusion and Discussion}










This study introduces a framework for promoting fairness in EV charger planning within a decentralized market. Using a bi-level optimization model, we first capture the competitive dynamics among providers via a Nash equilibrium. We then define socially optimal outcomes from both utilitarian and fairness-based perspectives. Finally, we design a strategic subsidy mechanism that steers the market's profit-driven behavior toward these desirable goals, offering a practical tool for regulators to balance efficiency with equitable public access.

Our findings demonstrate that by strategically adjusting unit capacity costs, the proposed subsidy effectively incentivizes investment in more distant stations. This redistribution of resources alleviates congestion at central locations and reduces access disparities across the network with only a modest sacrifice of overall market efficiency.

This study's limitations present several clear opportunities for future research. A key limitation of this study is its underlying definition of fairness, which prompts further research into the explicit trade-off with efficiency and the selection of an appropriate disagreement point for the NBS. In addition, there are many different perspectives to consider fairness, such as parity-based \cite{wang2026}, Max–Min \cite{ye2017}, and more. 
The current model's assumption of customer homogeneity can also be addressed by incorporating demographic data to consider equity and choice patterns. To summarize, future work should move beyond the simplified simulation by validating the model against a real-world dataset and exploring alternative fairness criteria.

Ultimately, our study provides policymakers with a quantitative method to bridge the gap between a profit-driven market and a more equitable, socially optimal system for EV infrastructure.

\bibliographystyle{IEEEtran}
\bibliography{bibliography.bib}

\end{document}